\documentclass[
preprint,
showpacs,
showkeys
]{revtex4-1}
\usepackage{amsthm}
\theoremstyle{plain}
\newtheorem*{theorem*}{Theorem}

\newtheorem*{definition*}{Definition}

\newtheorem*{lemma*}{Lemma}

\usepackage{braket}
\usepackage{delarray}
\usepackage{here}

\usepackage{txfonts}

\usepackage[dvipdfmx]{graphicx}
\usepackage{bm}
\usepackage{color}
\usepackage{ulem}

\newcommand{\be}{\begin{eqnarray}}
\newcommand{\ee}{\end{eqnarray}}
\newcommand{\ba}{\begin{array}}
\newcommand{\ea}{\end{array}}
\newcommand{\bmat}{\left(\begin{array}}
\newcommand{\emat}{\end{array}\right)}

\begin{document}
\title{
Bound on the distance between controlled quantum state \\ 
and target state under decoherence
}

\author{Kohei Kobayashi$^1$}

\affiliation{
$^1$Global Research Center for  Quantum Information Science, National Institute of Informatics,
 2-1-2 Hitotsubashi,  Chiyoda-ku, Tokyo 101-8340, Japan}

\begin{abstract} 

To implement quantum information technologies, carefully designed control for preparing a desired state plays a key role.
However, in realistic situation, the actual performance of those methodologies is severely limited by decoherence.
Therefore, it is important to evaluate how close we can  steer the controlled state to a desired target state under decoherence.
In this paper, we provide an upper bound of the distance between the two controlled quantum systems in the presence and absence of decoherence.
The bound quantifies the degree of achievement of the control for a given target state under decoherence, and 
can be straightforwardly calculated without solving any equation.
Moreover, the upper bound is applied to derive a theoretical limit of the probability for obtaining the target state under decoherence.

%\keywords{measurement-based feedback control, von Neumann entropy, stocahstic master equation, Sagawa-Ueda inequality}

\end{abstract}
\date{\today}
\maketitle

%%%%%%%%%%%%%%%%%%%%%%%%%%%%%%%%%%%%%%%%%%%%%%%%%%%%%%%%%%%%%%%%%%%%%%%%%%%
%%%%%%%%%%%%%%%%%%%%%%%%%%%%%%%%%%%%%%%%%%%%%%%%%%%%%%%%%%%%%%%%%%%%%%%%%%%
\section{Introduction}

Control of quantum dynamics is of great importance for preparing a desired target state in quantum science and technology.
The control of closed quantum system has been well formulated.
For instance, the open-loop control (OLC)  offers powerful means for efficiently
implementing a target state \cite{ol1, ol2, ol3, ol4} or adiabatic quantum computation  \cite{aqc1, aqc2, aqc3}.
In recent years, extending these studies to open quantum system, where the system interacts with an environments, are underway.

However, all quantum systems are subjected to environmental effects in realistic  situations.
The main obstacle for control of open quantum systems is decoherence, 
which is the loss of quantum information of a system caused by the interaction between the system and the external environment \cite{Nielsen}.   
Due to this, the actual control performance is sometimes far away from the ideal one.
Therefore, given a control system and decoherence, 
it is important to clarify and characterize how close one can steer the quantum state to the desired target state under decoherence.

For closed quantum systems, detailed analysis of control limit in various cases have already been investigated  
\cite{close1, close2, close3, close4, close5}.
Extending these, many studies have been extensively conducted to clarify the limit of control for open quantum systems in recent years
 \cite{open1, open2, open3, open4, open5, open6, open7, open8}. 
However, these results are limited to specific situations and require integration and eigenvalue calculations, 
which poses the problem of computational difficulties.

Motivated by this background, in this paper we present a control limit for a Markovian open quantum system 
towards the target state under decoherence.
More specifically, we derive an upper  bound of the norm distance between the state obeying the ideal evolution
 and the another one obeying the noisy evolution under decoherence. 
This bound quantitatively represents the degree of achievement of control under decoherence.
Also, it can be straightforwardly computed only by the information about decoherence and control time.
Thanks to this, there is no need to solve any equations for obtaining the bound.
In particular, the bound becomes tighter when the decoherence is small or the driving time is small even if the decoherence is large,
 as will be demonstrated in several examples.
Moreover, we use the result to derive a theoretical limit of the probability for obtaining the target state under decoherence,
and apply it to quantum algorithm.

\section{ Main result}

\subsection{Dynamics of controlled quantum system under decoherence}

We explain a time evolution of the controlled quantum system.
Consider the state of the quantum system $\ket{\psi(t)}$ initially prepared in state $\ket{\psi(0)}=\ket{\psi_0}$, 
which evolves according to the Schr${\rm \ddot{o}}$dinger equation:

\begin{eqnarray}
\label{SE}
\frac{d\ket{\psi(t)} }{dt} =-i\hat{H}(t)\ket{\psi(t)}, 
\end{eqnarray}
where  $\hat{H}(t)$ is the time-dependent control Hamiltonian.
We wish to drive the system to the target state $\ket{\psi_g}$ at some final time $t=T$ (we set $\hbar=1$ in the following).

We also consider another state $\hat{\rho}(t)$ which has the same initial state 
$\hat{\rho}(0)=\ket{\psi_0}\bra{\psi_0}$ and 
satisfies the Markovian master equation \cite{master}:

\begin{eqnarray}
\label{ME}
\frac{d \hat{\rho}(t) }{dt}  = -i[ \hat{H}(t), \hat{\rho}(t)] + \mathcal{D}[ \hat{M} ]\hat{\rho}(t), 
\end{eqnarray}
where $\hat{M}$ represents the decoherence process and
\begin{equation}
\label{Lindblad}
\mathcal{D}[\hat{M}] \hat{\rho} 
=  \frac{1}{2}\left( [ \hat{M}, \hat{\rho} \hat{M}^\dagger]+[\hat{M}\hat{\rho}, \hat{M}^\dagger]\right).
\end{equation}
Ideally, if  $\hat{M}=0$, Eq. (\ref{ME}) is identical to Eq. (\ref{SE}) and $\hat{\rho}(T)=\ket{\psi_g}\bra{\psi_g}$ is achieved.
In this paper, we define the distance between $\hat{\rho}(t)$ and $\ket{\psi(t) }\bra{\psi(t)}$
 for $t\in[0, T]$ as the Frobenius norm \cite{Nielsen}:
\begin{eqnarray}
\label{cost}
D(t)= \| \ket{\psi(t) }\bra{\psi(t)}- \hat{\rho}(t) \|_{\rm F},  
\end{eqnarray}
where $ D(t) \in [0, \sqrt{2}]$ and $\| \hat{A}\|_{\rm F}:=\sqrt{{\rm Tr}[\hat{A}^\dagger \hat{A}]}$.

In general, it is impossible to achieve $D(t)=0$ at $t=T>0$ under the decoherence term $\mathcal{D}[\hat{M}]$. 
If $\hat{\rho}(t)$ is orthogonal to $\ket{\psi(t) }\bra{\psi(t)}$, i.e., $F(\ket{\psi(t)}, \hat{\rho}(t)):=\langle \psi(t)| \hat{\rho}(t)| \psi(t)\rangle=0$, 
$D(t)$ takes the maximum value $\sqrt{2}$.
Hence,  $D(t)$ can be regarded as the cost function between the controlled two quantum 
states in the absence and presence of decoherence.
Under these settings,  how much $D(t)$ increases at most is of great interest.

\subsection{Main result}

The main result of this paper is to present an upper bound
of the cost function  $D(t)$ in an explicit form.
The proof of this result is provided in Appendix A.

{\it Theorem 1.}  The cost function (\ref{cost})  has the following upper bound:

\begin{equation}  
\label{upperbound}
D(T) \leq   \delta:=
\left\{
\begin{array}{l@{\quad}l}
\sqrt{2}\beta T     & ( \hat{M}^\dagger \neq\hat{M} ),\\
\beta T    & (\hat{M}^\dagger =\hat{M}),
\end{array}
\right.
% \cases{  
%  &   \  $(  )$,      \\
%\beta T &  \     $(  )$, }
\end{equation}
where $\beta:=(1/T)\int^T_0 \| \hat{M} \|^2_{\rm F}dt$.
%{\bf $( \hat{M}^\dagger \neq\hat{M} )$},

If $T=0$ or $\hat{M}=0$, $\delta=0$ holds, otherwise $\delta>0$;  that is,  $\delta$ gives a fundamental limit on how close the controlled quantum state can be steered to a target state under decoherence.
 Below we list some notable facts regarding this inequality:

(i) When the system is subjected to multiple decoherence channels, $\beta$ can be straightforward extended as follows:
\begin{eqnarray}  
\beta = \frac{1}{T}  \int^T_0\sum_j  \|\hat{M}_j \|^2_{\rm F}dt.
\end{eqnarray}

(ii)  $\delta$ can be calculated by the information about the parameters $\hat{M}$ and $T$, which enables us not to solve any dynamical equations.

(iii) Here let us consider an important extension of (5); suppose that 
$\hat{H}=0$ and $\ket{\psi(t)}$ is fixed in the initial state $\ket{\psi_0}$ in Eq. (1),
and $\hat{\rho}(t)$ evolves ``undesirable'' final state satisfying 
$D(T)=\|\ket{\psi_0}\bra{\psi_0}-\hat{\rho}(T)\|_{\rm F}$ at $t=T$ under decoherence 
(we note that the final state here is undesirable, although we have assumed that $\ket{\psi(t)}$ reaches the desired target state at $t=T$ above). 
In this case, the inequality  (\ref{upperbound}) can be seen as the quantum speed limit (QSL),
which is the lower bound of the evolution time for a quantum system from an initial state to a final state \cite{MT, ML}:

\begin{equation}
\label{qsl}  
T \geq T_\delta:=\left\{
\begin{array}{l@{\quad}l}
\frac{D(T)}{ \sqrt{2}\beta}     & ( \hat{M}^\dagger \neq\hat{M} ),\\
\frac{D(T)}{ \beta}   & (\hat{M}^\dagger =\hat{M}).
\end{array}
\right.
%\cases{
%, & \   $( )$,    \\
%, & \  $( \hat{M}^\dagger =\hat{M} )$} .
\end{equation}
This $T_\delta$ is our QSL, giving a lower bound on the evolution time for the state $\hat{\rho}(t)$ to evolve from 
$\hat{\rho}(0)=\ket{\psi_0}\bra{\psi_0}$ to any final state $\hat{\rho}(T)$ satisfying a given value of $D(T)$.

Importantly, $T_\delta$ can be used as a tool for quantifying how close a quantum state can stay around its initial state under decoherence.
Let us consider the situation where an initial state $\hat{\rho}(0)=\ket{\psi_0}\bra{\psi_0}$ is ideal and a value of $D(T)$ are given. 
This means that we are given a region $\mathcal{R}_D(\ket{\psi_0})$, which is the set of all states whose distance from $\hat{\rho}(0)$
is less than $D(T)$, that is, $D(T)$ can be interpreted as the radius of a circle region $\mathcal{R}_D(\ket{\psi_0})$.
Then, the transition time $T$ for evolving $D(0)=0$ to $D(T)$ has the meaning of the escape time that 
the state first exits from $\mathcal{R}_D(\ket{\psi_0})$ due to decoherence.
Therefore, $T_\delta$ corresponds to the condition for the state $\hat{\rho}(t)$ to remain in a given region $\mathcal{R}_D(\ket{\psi_0})$.

Of course, $T_\delta$ can be also explicitly calculated in terms of the parameters $\hat{M}$, once the value of $D(T)$ is specified. 
Moreover, as will be shown later, $T_\delta$ is always tighter than a typical bound given in \cite{Campo} in particular setup. 

In what follows, we will examine the effectiveness of $\delta$ through some examples.

\section{Example}

\subsection{Qubit}

We first consider a single qubit system composed of the excited state $\ket{0}=[1, 0]^\top$ and the ground state $\ket{1}=[0, 1]^\top$.
We aim to drive the state from $\ket{\psi_0}=\ket{1}$ to $\ket{\psi_g}=\ket{0}$ under the setting:
\begin{eqnarray}
\hat{H}=\Omega \hat{S}_y, \ \ \hat{M}=\sqrt{\gamma} \hat{S}_-,
\end{eqnarray}
where we defined $\hat{S}_x=\ket{0}\bra{1}+\ket{1}\bra{0}$, $\hat{S}_y=-i(\ket{0}\bra{1}-\ket{1}\bra{0})$, 
and $\hat{S}_z=\ket{0}\bra{0}-\ket{1}\bra{1}$
and $\hat{S}_-=\ket{1}\bra{0}$.
$\hat{H}$ represents the control Hamiltonian with the time-independent driving frequency $\Omega>0$. 
$\hat{M}$ represents the decay process corresponding the spontaneous emission, with the decay rate $\gamma>0$.
In the ideal case, i.e., $\gamma=0$, by solving the equation $d\ket{\psi(t)}/dt=-i\Omega \hat{S}_y\ket{\psi(t)}$, 
we obtain the exact time required for the state transition $\ket{1}\to\ket{0}$ as $T=\pi/(2\Omega)$.
 Hence, the upper bound for the system driven by Eq. (8) is given by
$\delta=\pi \gamma/(\sqrt{2}\Omega)$.

We here focus on the difference $\Delta$ between $\delta$ and the actual distance $D(T)$:
\begin{eqnarray}
 \Delta=\delta-D(T),
\end{eqnarray}
where $\Delta \in [0, \sqrt{2}]$. We can say that  the closer $\Delta$ is to zero, the lesser is the effect of the interaction with the environment.

Figure 1(a) illustrates the simulated values of $\Delta$ as a function of $\gamma$, for $\Omega=1$, $3$, and $10$.
 Note that the last point when $\Omega=1$ is meaningless, because it is greater than $\sqrt{2}$.
This means that $\gamma=1$ is too large for the upper bound $\delta$ to work meaningfully, 
regarding to the control time achieved by the driving Hamiltonian with $\Omega=1$.
Meanwhile, as $\Omega$ becomes large, $\Delta$ becomes small for all $\gamma$.
Thereby,  for the rapid state transition, $\delta$ gives a better estimate on the achievement of state preparation even if $\gamma$ is large.

Here it is also worth comparing $\delta$ with the exact distance $D(T)$ to see the tightness $\delta$. 
For this purpose, here we set $\hat{H}=0$, $\hat{M}=\sqrt{\gamma}\hat{S}_z$ and choose the initial state 
as  the superposition $\ket{\psi_0}=\ket{+}:=(\ket{0}+\ket{1})/\sqrt{2}$. In order to obtain the exact value of $D(T)$, it is needed to know the $x$, $y$, and $z$ components of $\hat{\rho}(t)$, which are defined by 
\begin{equation}
\hat{\rho}(t)=\frac{1}{2}\left(\hat{I}+x(t)\hat{S}_x+y(t)\hat{S}_y+ z(t)\hat{S}_z\right).
\end{equation}
Then, by solving the master equation $d\hat{\rho}(t)/dt=\mathcal{D}[\hat{M}]\hat{\rho}(t)$, we find
 \begin{eqnarray}
&& x(t)=e^{-2\gamma t}, \\
&&  y(t)=z(t)=0.
\end{eqnarray}

As a result, $D(T)$ and $\delta$ can be computed as follows:

\begin{eqnarray}
D(T)=\frac{1-e^{-2\gamma T} }{\sqrt{2}}, \ \ \ \  \delta=\sqrt{2}\gamma T.
\end{eqnarray}

Figure 1(b) shows the plots of $D(T)$ and $\delta$ as a function of $\gamma T$.
We find that  both  $D(T)$ and $\delta$ are close to zero when $\gamma T$ is small, which is reasonable 
because the state $\hat{\rho}(t)$ does not move away from the initial state.
However, $\Delta$ diverges, as $\gamma T$ increases.
This fact suggests that we should use $\delta$ only when $\gamma T$ is small.

Finally, let us see the  tightness of the inequality (\ref{upperbound})  as a QSL by comparing 
with another bound; in Ref \cite{Campo}, an explicit QSL was derived:
\begin{eqnarray}
T \geq T_{\rm DC}:=\frac{ 1- F(\ket{\psi_0}, \hat{\rho}(T))   }{\| \mathcal{D}^\dagger [\hat{M}] \ket{\psi_0}\bra{\psi_0} \|_{\rm F} },
\end{eqnarray}
where $\mathcal{D}^\dagger [ \hat{M}] \hat{\rho}=\hat{M}^\dagger \hat{\rho} \hat{M}-\hat{M}^\dagger \hat{M}\hat{\rho}/2-\hat{\rho} \hat{M}^\dagger \hat{M}/2$.
Note that $T_{\rm DC}$ is explicitly represented in terms of $(\ket{\psi_0}, \hat{M}, \hat{\rho}(T) )$, 
which enables us to compare our bound $T_{\delta}$ with $T_{\rm DC}$.
In particular, in the setup $(\ket{\psi_0}, \hat{H}, \hat{M})=(\ket{+}, 0, \sqrt{\gamma}\hat{S}_z)$,
$T_{\delta}$ and $T_{\rm DC}$ are calculated as follows:
\begin{eqnarray}
T_{\delta}=\frac{1-e^{-2\gamma T}}{2\gamma}, \ \ \ \  T_{\rm DC}:=\frac{1-e^{-2\gamma T} }{2\sqrt{2}\gamma}.
\end{eqnarray}
Thus, $T_\delta$ is always tighter than $T_{\rm DC}$ in this case.
Therefore, we can say that $T_\delta$ can offer higher performance in practical use, even under little information.

\begin{figure}[tb]
\includegraphics[width=14cm]{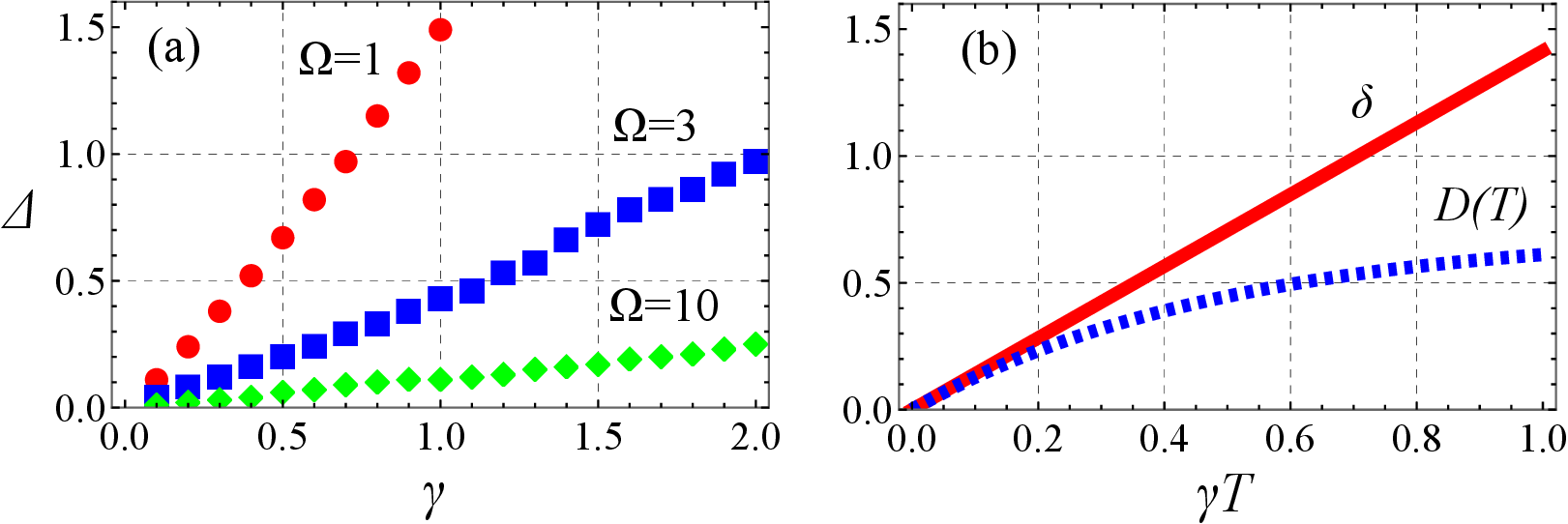}
\caption{  Plots of $\Delta$ as a function of $\gamma$
when $\Omega=1$, $3$, and $10$.   These plots are computed via numerical simulation.
(b) Comparison of the exact distance $D(T)$ and its upper bound $\delta$ as a function of $\gamma T$.}
\end{figure}

\subsection{Two qubits}

Next we study the two-qubit system under decoherence.
Let us focus on the SWAP operation realized by the SWAP gate:

\begin{eqnarray}
 \hat{U}_{\rm SWAP}  =
 \left[  \begin{array}{cccc}
 1 & 0 & 0 & 0   \\ 
0 & 0 & 1 & 0   \\
0& 1 & 0 &0   \\
0 & 0 & 0& 1\end{array}
\right],  
\end{eqnarray}

which exchanges the state of qubits 1 and 2 and plays an important role in quantum information processing 
such as a quantum Fourier transform \cite{Nielsen}.
It is known that $\hat{U}_{\rm SWAP} $ is implemented by the time-independent  Hamiltonian
\begin{eqnarray}
\hat{H}=\frac{\Omega}{2}\left( \hat{S}_x\otimes \hat{S}_x  + \hat{S}_y\otimes \hat{S}_y +\hat{S}_z\otimes \hat{S}_z \right).
\end{eqnarray}
We choose the initial state as the product state $\ket{\psi_0}=\ket{+}\otimes \ket{0}$, and then the final target state is given by
\begin{eqnarray}
\ket{\psi_g}= \hat{U}_{\rm SWAP} \ket{\psi_0}=\ket{0}\otimes \ket{+},
\end{eqnarray}
which takes the transition time $T=\pi/(2\Omega)$.
Furthermore, we here consider the two types of decoherences;
amplitude damping $\hat{M}_{\rm AD}$ and phase damping  $\hat{M}_{\rm PD}$ \cite{open7}:
\begin{eqnarray}
\hat{M}_{\rm AD} &=& \frac{\sqrt{\gamma}}{2}( \hat{S}_-\otimes \hat{I} +\hat{I} \otimes \hat{S}_-), \\
\hat{M}_{\rm PD} &=& \frac{\sqrt{\gamma}}{2} (\hat{S}_z\otimes \hat{I} +\hat{I} \otimes\hat{ S}_z).
\end{eqnarray}
$\hat{M}_{\rm AD}$ represents the energy decay process which acts on the two atoms spontaneously.
$\hat{M}_{\rm PD}$ represents the loss of information on the phase of the whole system.
Due to the fact that $\| \hat{M}_{\rm AD}\|^2_{\rm F}=\| \hat{M}_{\rm PD}\|^2_{\rm F}=\gamma$
 and $\hat{M}^\dagger_{\rm AD}\neq \hat{M}_{\rm AD}$ 
 but $\hat{M}^\dagger_{\rm PD}=\hat{M}_{\rm PD}$,
the upper bound $\delta$ for  $\hat{M}_{\rm AD}$ and  $\hat{M}_{\rm PD}$ are calculated as  follows:
\begin{eqnarray}
\delta(\hat{M}_{\rm AD})= \frac{\gamma \pi}{ \sqrt{2} \Omega}, \ \ \ \ \delta(\hat{M}_{\rm PD})=\frac{\gamma \pi}{2\Omega}.
\end{eqnarray}
Figure 2 (a) and (b) show $\Delta=\delta-D(T)$ as a function of $\gamma$ for $\hat{M}_{\rm AD}$ and $\hat{M}_{\rm PD}$, respectively.
Under the same value of $\Omega$, for a given $\gamma$, $\Delta$ for $\hat{M}_{\rm PD}$ is always smaller than that for $\hat{M}_{\rm AD}$.
This result indicates that $\delta$ becomes tighter when the  decoherence operator is Hermitian in this case, but 
whether this holds in general or not is unclear.

\begin{figure}[tb]
\includegraphics[width=14cm]{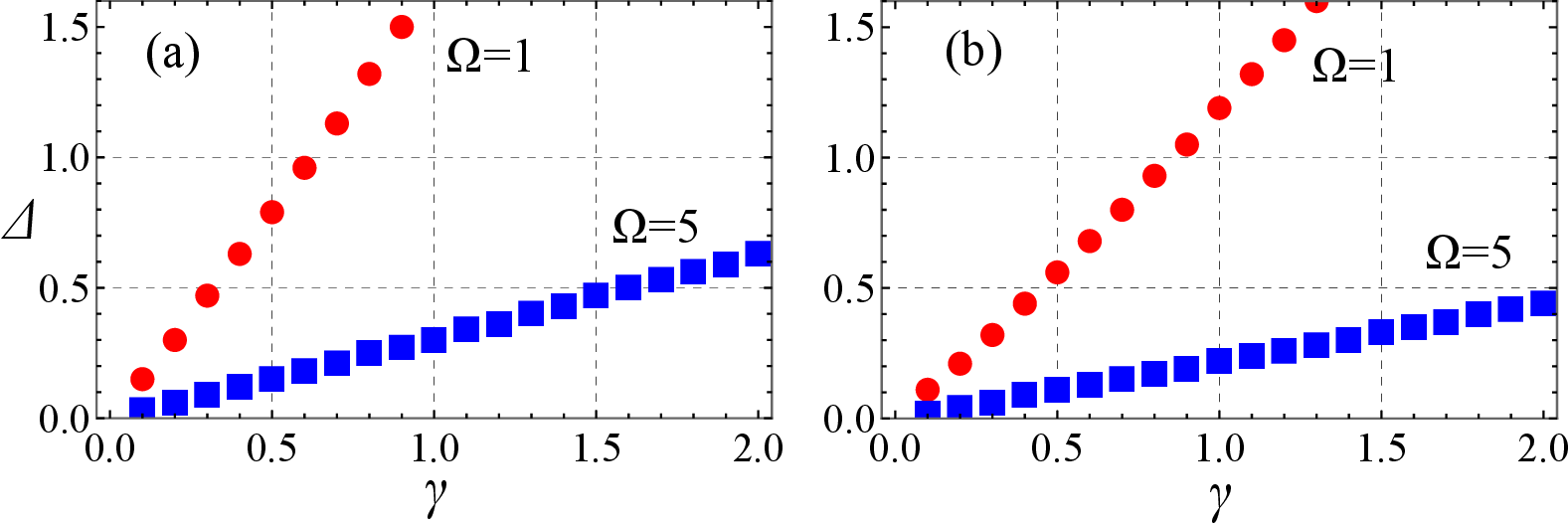}
\caption{  Plots of $\Delta$ for (a) $\hat{M}_{\rm AD}$ and (b) $\hat{M}_{\rm PD}$, as a function of $\gamma$ 
when $\Omega=1$ and $5$. }
\end{figure}

\subsection{Bound on probability for obtaining the target state }

Lastly, we turn our attention to the probability for obtaining the target state in the presence of decoherence.
More specifically, we examine how much the probability for obtaining the target state decreases from $1$ compared than the ideal case $\hat{M}=0$.

We here decompose the final state $\hat{\rho}(T)$ as follows:
\begin{eqnarray}
\hat{\rho}(T) =\sum^n_{i=1}p_i\ket{i}\bra{i}.
\end{eqnarray}
 We assume that the desired target state is the $m$th eigenstate of $\rho(T)$, $\ket{\psi_g}=\ket{m}$, which is obtained via projective measurement.
What is of our interest here is at least to what extent the probability of obtaining the target state is guaranteed under noise.
Therefore, in the above setting, we derive a theoretical lower bound of $p_m$ by using  $\delta$.

We consider  $D^2(T)\leq \delta^2$ and 
\begin{eqnarray}
\label{D}
D^2(T) &=&  \| \ket{\psi_g}\bra{\psi_g} - \hat{\rho}(T) \|^2_{\rm F}   \nonumber  \\
&=& 1-2\langle m | \hat{\rho}(T)| m \rangle + {\rm Tr}\left[\hat{\rho}^2(T) \right]   \nonumber  \\
&=& 1-2\langle m| \left(\sum^n_{i=1}p_i\ket{i}\bra{i}\right) |m\rangle + \sum^n_{i=1}p^2_i    \nonumber \\
&\geq& 1-2p_m +p_m^2  \nonumber \\
&=& (1-p_m)^2.
\end{eqnarray}

Therefore, we obtain the following inequality:
\begin{eqnarray}
\label{probbound}
p_m \geq 1-\delta.
\end{eqnarray}

 When the inequality $\beta<1/T$ holds,  the righthand side of Eq. (\ref{probbound}) works as a meaningful bound.
Therefore, if $T$ is large, the probability of obtaining the target state under projective measurement
 is estimated to be small, even if the decoherence is small. 
%Therefore, if $T$ is large, even for the small decoherence, the probability for obtaining the target state is estimated.
%In general, it may seem difficult to make the lower bound make sense when the computing time $T$ is large.
%Therefore, the decoherence must be highly suppressed when $T$ is large.

For instance, we apply this result to a quantum algorithm to locate the state $\ket{m}$ in an unsorted database set of normalized orthogonal states
$\{\ket{i}, i=1,\cdots,m,\cdots n \}$.
It is known that this algorithm has a computational complexity $\mathcal{O}(\sqrt{n})$ as that of Grover's algorithm.
To make the lower bound independent of the system size,
the decoherence term should be at least suppressed as small as
 \begin{eqnarray}
\beta =\mathcal{O}(n^{-\frac{1}{2}}).
\end{eqnarray}
Thus, any decoherence processes, if its strength is less than the inverse of the computing time,
the target state may be obtained in a constant-order.

\section{Conclusion}

In this paper, we have derived the upper bound $\delta$ of the distance 
between the two controlled quantum states in the absence and presence of decoherence.
$\delta$ shows better performance when the strength of the decoherence is small or the operator representing the decoherence is Hermitian.
Also it can be easily computed without solving any equation, which enables us to apply this result to wide class of quantum control.
The equality condition of obtained inequality $D(T)\leq \delta$ is unclear. 
Thus, finding the class of control setup for achieving $\delta$ is an interesting future work.
Moreover, based on $\delta$, we presented a theoretical lower bound of the probability for the desired target state 
that is obtained via projective measurement
under decoherence.
This lower bound theoretically indicates at least how well the target state can be controlled under decoherence.
Finally, we emphasize that any assumption is not imposed on time evolution.
Taking into account a special schedule for each quantum protocol may improve the present result.

This work was supported by MEXT Quantum Leap Flagship Program Grant JPMXS0120351339.

\appendix

\section{Proof of the theorem 1}

We begin with taking the time derivative of $\hat{\zeta}(t):=\ket{\psi(t)}\bra{\psi(t)}- \hat{\rho}(t)$:

\begin{eqnarray}
\label{derive1}
\frac{d \hat{\zeta}(t) }{dt} &=& -i[ \hat{H}(t), \ket{\psi(t)}\bra{\psi(t)}]-\left(-i[ \hat{H}(t), \hat{\rho}(t)] 
+ \mathcal{D}[\hat{M}] \hat{\rho}(t) \right) \nonumber \\ 
&=&   -i[ \hat{H}(t), \hat{\zeta}(t)] -\mathcal{D}[ \hat{M}] \hat{\rho}(t).
\end{eqnarray}

Next we consider

\begin{eqnarray}
\label{derive2}
\frac{dD^2(t) }{dt}  &=&  2 {\rm Tr}\left[ \hat{\zeta}(t) \frac{d \hat{\zeta}(t) }{dt} \right]  \nonumber \\
&=& 2 {\rm Tr}\left[ \hat{\zeta}(t) \left( -i[ \hat{H}(t), \hat{\zeta}(t)] - \mathcal{D}[ \hat{M}] \hat{\rho}(t)\right)  \right]   \nonumber \\ 
&=&  - 2 {\rm Tr}\left[ \hat{\zeta}(t) \mathcal{D}[ \hat{M}] \hat{\rho}(t)  \right]  
\end{eqnarray}

In order to bound the second term of the righthand side of (\ref{derive2}), 
we here introduce the following inequality \cite{Bottcher}:
\begin{eqnarray}
\|[ \hat{X}, \hat{Y}]\|_{\rm F} \leq \sqrt{2}\| \hat{X}\|_{\rm F} \| \hat{Y}\|_{\rm F},
\end{eqnarray}
where $\hat{X}$ and $\hat{Y}$ are any matrices.
Using this inequality and (\ref{Lindblad}) for non-Hermitian operator $\hat{M}$, 
we calculate the lower bound as follows:
\begin{eqnarray}
\label{derive5}
-2 {\rm Tr}\left[ \hat{\zeta}(t) \mathcal{D}[ \hat{M} ]\hat{\rho}(t) \right] 
&=&-  {\rm Tr}\left[ \hat{\zeta}(t) \left( [ \hat{M}, \hat{\rho}(t) \hat{M}^\dagger] 
+[\hat{M} \hat{\rho}(t), \hat{M}^\dagger] \right)\right]    \nonumber \\
&\leq& \left| {\rm Tr}\left[ \hat{\zeta}(t) \left( [ \hat{M}, \hat{\rho}(t) \hat{M}^\dagger] 
+[ \hat{M} \hat{\rho}(t), \hat{M}^\dagger] \right)\right] \right| \nonumber \\
&\leq&D(t) \| [ \hat{M}, \hat{\rho}(t) \hat{M}^\dagger]+ [ \hat{M} \hat{\rho}(t), \hat{M}^\dagger]  \|_{\rm F} \nonumber \\
&\leq&D(t) \left( \| [ \hat{M}, \hat{\rho}(t) \hat{M}^\dagger]\|_{\rm F} +  \|[ \hat{M}\hat{\rho}(t), \hat{M}^\dagger]  \|_{\rm F} \right)  \nonumber \\
&\leq&\sqrt{2} \left(  \| \hat{M}\|_{\rm F}  \| \hat{\rho}(t) \hat{M}^\dagger\|_{\rm F}  
+ \| \hat{M} \hat{\rho}(t)\|_{\rm  F} \| \hat{M}^\dagger \|_{\rm F}  \right)  D(t)  \nonumber \\
&\leq&2\sqrt{2}\| \hat{M}\|^2_{\rm F}\| \hat{\rho}(t)\|_{\rm  F}  D(t) \nonumber \\
&\leq&2\sqrt{2}\| \hat{M} \|^2_{\rm F} D(t). 
\end{eqnarray}
We used the Schwarz's inequality $|{\rm Tr}[ \hat{A}\hat{B}]| \leq \| \hat{A}\|_{\rm F} \|\hat{B}\|_{\rm F}$ and $\| \hat{\rho}\|_{\rm  F}\leq 1$.

On the other hand, we find
\begin{eqnarray}
\label{derive6}
\frac{dD^2(t)  }{dt} =2D(t)\frac{dD(t)}{dt}.
\end{eqnarray}

Combining  these relations,  we have
\begin{eqnarray}
\label{derive6}
\frac{dD(t)  }{dt} \leq  \sqrt{2}\| \hat{M}\|^2_{\rm F}.
\end{eqnarray}
By integrating both sides of Eq. (\ref{derive6}) from $0$ to $T$, 
we end up with the upper  bound:
\begin{eqnarray}
\label{deriv7}
D(T) \leq \delta:=\sqrt{2}\beta T,
\end{eqnarray}
where we defined $\beta= (1/T)\int^T_0 \| \hat{M} \|^2_{\rm F}dt$.

Also, if $\hat{M}$ is Hermitian, $\mathcal{D}[\hat{M}] \hat{\rho}= [[ \hat{M}, \hat{\rho}], \hat{M}]/2$. 
Hence,  Eq. (\ref{derive5}) can be tightened as follows.
\begin{eqnarray}
-2 {\rm Tr}\left[ \hat{\zeta}(t) \mathcal{D}[ \hat{M} ]\hat{\rho}(t) \right] 
&=&  - {\rm Tr}\left[ \hat{\zeta}(t) [[\hat{M}, \hat{\rho}(t)], \hat{M}]  \right]    \nonumber \\
&\leq& \left| {\rm Tr}\left[ \hat{\zeta}(t) [[ \hat{M}, \hat{\rho}(t)], \hat{M}]  \right] \right| \nonumber \\
&\leq&D(t) \| [[ \hat{M}, \hat{\rho}(t)], \hat{M}]  \|_{\rm F} \nonumber \\
&\leq& \sqrt{2} D(t) \|[ \hat{M}, \hat{\rho}(t)] \|_{\rm F}  \| \hat{M}\|_{\rm F} \nonumber \\
&\leq&2\| \hat{M}\|^2_{\rm F} \| \hat{\rho}(t)\| D(t)  \nonumber \\
&\leq&2\|\hat{M} \|^2_{\rm F}D(t).
\end{eqnarray}

In the same manner, we have $D(T) \leq \delta:=\beta T$.

%%%%%%%%%%%%%%%%%%%%%%%%%%%%%%%%%%%%%%%%%%%%%%%%%%%%%%%%%%%%%%%%%%%%%%%%%%%%

\end{document}